  \documentclass[prb,showpacs,twocolumn,amssymb,amsmath,nobibnotes]{revtex4}

\usepackage{graphicx,color}
\usepackage{dcolumn}
\usepackage{bm}

\begin{document}

\title{Mn-doped Thiolated Au$_{25}$ Nanoclusters: Atomic Configuration, Magnetic Properties, and A Possible High-performance Spin Filter}

\author{M. Zhou$^{1}$, Y. Q. Cai$^{1}$, M. G. Zeng,$^{1}$ C. Zhang$^{1,2,*}$, and Y. P. Feng$^{1,\dag}$}
\affiliation{$^{1}$ Department of Physics,National University of Singapore, 2 Science Drive 3,Singapore, 117542\\
$^{2}$Department of Physics and Chemistry, National University of Singapore, 3 Science Drive 3, Singapore, 117543\\
$^{*}$ phyzc@nus.edu.sg, $^{\dag}$ phyfyp@nus.edu.sg}

\begin{abstract}
We report an \emph{ab inito} investigation on the ground-state atomic configuration, electronic structures, magnetic and spin-dependent transport properties of Mn-doped Au$_{25}$ nanoclusters protected by thiolate. It is found that the most stable dopant sites are near surfaces, rather than the center positions of the nanoparticles. Transport calculations show that high-performance spin filters can be achieved by sandwiching these doped clusters between two nonmagnetic Au electrodes. The nearly perfect spin filtering originates from localized magnetic moments of these clusters that are well protected by ligands from the presence of electrodes.
\end{abstract}

\maketitle

Thanks to recent advances in experimental techniques such as wet chemical synthesis and atomical control of nanoparticles,\cite{syth} there has been a surge in research pertaining to effects of ligand protection on properties of Au clusters or nanoparticles. Thiolate-protected Au clusters have been shown to have unique properties that imply great potential in practical applications such as nanocatalytsts, sensors, biological markers, or fluorescence.\cite{catalysis, sensors, flor} Theoretical investigations suggested that the stability of these ligand-protected clusters is closely related to the electron shell closing, leading to magic numbers, in analogy with the case of single atoms.\cite{PNAS} Au$_{25}$(SR)$_{18}$$^-$ (SR = organothiolate group) is a well-studied typical example of these thiolate-protected Au clusters, which constitutes a compact icosahedral Au$_{13}$ core protected by six [(SR)$_3$Au$_2$] motifs, and exhibits extremely high stability with a large optical gap of 1.3 eV.\cite{Au25_1, Au25_2} Recent experimental and theoretical studies have shown that electronic and magnetic properties of Au$_{25}$(SR)$_{18}$$^-$ can be efficiently controlled and tuned by chemical doping of foreign elements. \cite{Au24Pd_ex,CuAgCd_dope,mod3,mod4} Versatile electronic, optical, and magnetic properties of Au$_{25}$(SR)$_{18}$$^-$ clusters doped by different atoms make these clusters promising candidates for various kinds of applications in molecular electronic, optical, and spintronic devices.

We are particularly interested in the doping-induced magnetic properties of these ligand-protected Au clusters in the context of molecular magnet or magnetic superatom, \cite{mole_mag1, mole_mag2} and their possible applications in molecular spintronics. Here, by first principles approaches, we investigate ground-state geometrical, electronic, magnetic, and spin-dependent transport properties of Mn-doped Au$_{25}$(SR)$_{18}$$^-$ clusters. We show that the most favorable doping position of the Mn atom is the surface site, different from previously suggested icosahedral-center doping, \cite{mod3,mod4} and these Mn-doped clusters may be used as nearly perfect spin filters. The spin-filtering we observed originates from the ligand-protected magnetic moments of clusters, which doesn't require the symmetry breaking of two spin channels in electrodes that is normally indispensable in usual spintronic devices. \cite{Fe_MgO_Fe}

Our first-principles electronic structure calculations are performed using projector augmented wave formalism of DFT through VASP package.\cite{VASP} The calculations are performed with a plane wave basis (30 Ry for the kinetic energy cutoff), and the generalized gradient approximation (GGA) in Perdew-Burke-Ernzerhof (PBE) format\cite{PBE} was included. To optimize the structure and magnetic moment, the clusters are put in a (25$\times$25$\times$25 \textrm{\AA}$^3$) cubic box, a single $\Gamma$-point for sampling, and the force convergence criterion is set to 0.01 eV/\textrm{\AA}$^3$. The transport calculations are based on nonequilibrium Green's function as implemented in the ATK package.\cite{ATK1,ATK2} The cut-off energy is 150 Ry and a Monkhorst-Pack k-mesh 0f 1$\times$1$\times$300 is used, with a double-$\zeta$  polarized basis set.

\begin{table}[h]
\caption{\label{tab:test}Relative energies, geometrical and electronic properties of Au$_{24}$Mn(SH)$_{18}$ for dopant at C, S and L sites, corresponding to configurations in Fig 1(a), (b) and (c), respectively. Relative energies ($\Delta$E) are taken the energy of S-site doping as reference. M$_{total}$ (M$_{Mn}$) is the total magnetic moment (single Mn dopant) of the clusters, and d$_{Mn-Au}$ (d$_{Au-Au}$) denotes the average distance for Mn-Au (Au-Au) bonds.}
\begin{tabular}{cccccc}
  \hline
  \hline
  Site & $\Delta$E (eV)	& M$_{total}$ ($\mu$B) & M$_{Mn}$ ($\mu$B) & d$_{Mn-Au}$ (\AA) & d$_{Au-Au}$ (\AA)\\
  \hline
  C & 0.42 & 5 & 4.07	& 2.81 & 2.98 \\
  \hline
  S & 0 & 5	& 4.21 & 2.70 & 2.89 \\
  \hline
  L & 0.24 & 5 & 4.15 & 2.66 & 2.93\\
  \hline
 \end{tabular}
 \label{t1}
\end{table}

\begin{figure}
  \includegraphics[width=8.0cm]{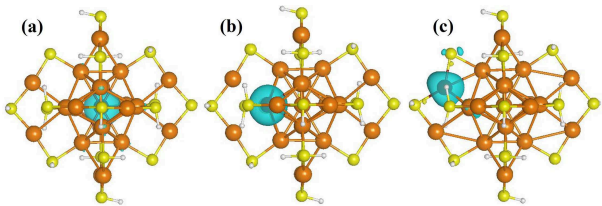}
  \caption{(Color online) Optimized structures for Au$_{24}$Mn(SH)$_{18}$ clusters. (a) Mn doped at the icosahedral Au$_{12}$ center sites, (b) at the surface site and (c) at the ligand.Au: orange (big), Mn: black (big), S: yellow (medium), H: white (small). Superimposed are the spin densities for the three cases, defined as the difference between spin-up and spin-down electron densities,$\rho$$\uparrow$-$\rho$$\uparrow$.}
\end{figure}

In Fig. 1, we show the optimized atomic configurations for thiolate protected Au$_{25}$ clusters doped by a Mn atom at different positions (Au$_{24}$Mn(SH)$_{18}$). The organothiolate group SR is replaced by SH for computational simplicity.\cite{mod4,JPCL} Three high symmetric doping sites were considered: The Au$_{13}$ core center site (C), the core surface site (S), and the ligand site (L). Note that for this system, the closed-shell configuration corresponds to the charge neutral state.\cite{PNAS} Our calculations show that the S site doping is the lowest-energy configuration. The L site doping is 0.24 eV higher, and the previously suggested C site doping turns out to be the most unstable one (0.42 eV higher than the lowest-energy state (S site)), which may be understood by the fact that the C site doping causes the largest expansion of the cluster as suggested by average bond lengths of Mn-Au (d$_{Mn-Au}$) and Au-Au (d$_{Au-Au}$) listed in Table I.
We also considered other ligand groups such as SCH$_3$ and SC$_2$H$_4$Ph. It is found that for all cases, the site near the cluster surface is more energetically favorable than the center site.

According to our calculations, the undoped neutral Au$_{25}$(SH)$_{18}$ cluster has one \emph{uB} of magnetic moment due to the unpaired spin, and the spin density defined as the difference between spin-up and spin-down electron densities is spread over the whole cluster, similar to the case of neutral Au$_{25}$(SR)$_{18}$ cluster. \cite{neutral}  The Mn doped cluster, Au$_{24}$Mn(SH)$_{18}$, has a large magnetic moment of 5 \emph{uB} for all three doping configurations. The spin densities are highly localized on the Mn atom as shown in Fig. 1 for doped cases. In order to see clearly the effects of ligand protection on magnetic properties of the doped cluster, we removed the ligand shell (6 (SH)3Au2 motifs), and investigated the Mn-doped icosahedral Au$_{12}$ cluster (the core). We found that the magnetic moment in this case is greatly reduced to about 1 \emph{uB}.

\begin{figure}
  \includegraphics[width=9cm]{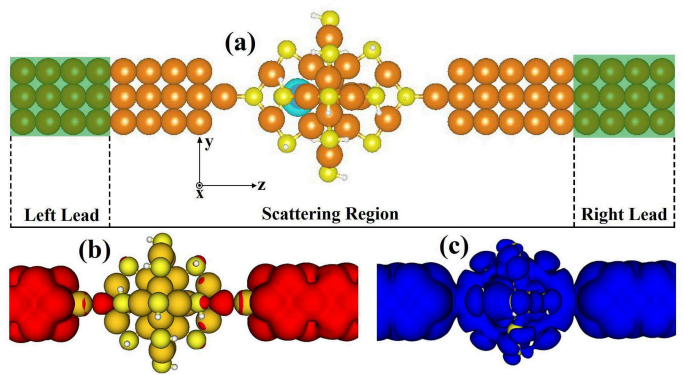}
  \caption{(Color online) (a) A schematic device model of most stable Au$_{24}$Mn(SH)$_{18}$ cluster bridging two gold leads. The scattering region includes the cluster and several Au surface layers. Superimposed are the spin densities for the total system. (b) and (c) show surfaces of the constant spin-resolved local DOS evaluated at the Fermi level. (b): $\alpha$-spin, (c): $\beta$-spin.}
\end{figure}

We then consider the possible application of the Au$_{24}$Mn(SH)$_{18}$ cluster in molecular spintronics. A schematic view of a molecular device consisting of a Mn-doped thiolated Au cluster and two Au electrodes is shown in Fig. 2a. An additional Au atom is added to each end of the leads to simulate atomically sharp tips.\cite{??} The distances between the cluster and gold leads were fully optimized. Due to the protection of the ligand shell, two electrodes have little effects on the magnetic state localized on the Mn atom as suggested by the spin density plot in Fig. 2a. In current work, we focus on the equilibrium electron transport.

\begin{figure}
  \includegraphics[width=9cm]{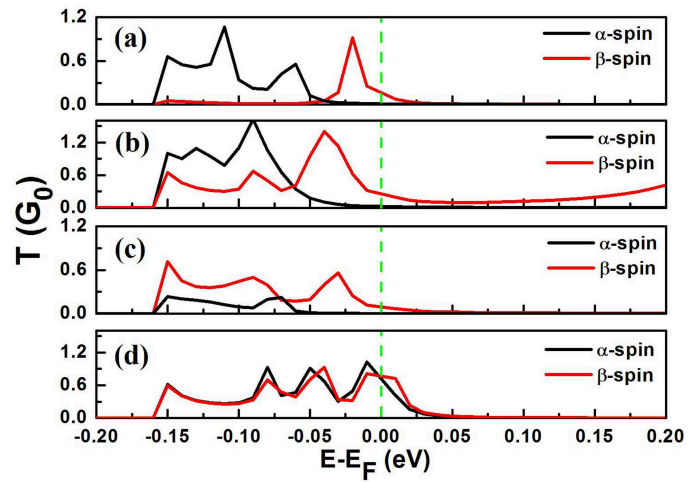}\\
  \caption{(Color online) The spin-dependent electron transmission at zero bias for the devices. (a), (b) and (c) correspond to Mn doped at S, C and L sites, respectively. (see Fig.1 for details) (d) shows the electron transmission for the undoped case. Fermi energy is set to zero (vertical dashed line).}
\end{figure}

Equilibrium spin-dependent conductance spectra for S, C, and L doping configurations are shown in Fig. 3a, b, and c, respectively. For all doping cases, around the Fermi energy, the spin-up channel (majority spin) is essentially close, while the spin-down chanel (minority spin) shows significant conductances. We plot isosurfaces of tranport eigen channels at Fermi energy for two spin channels for the case of S site doping in Fig. 3d, from where we can see that for the spin-down channel, the cluster is disconnected from two electrodes, and for the spin-up channel, the cluster is well connected because of the localized magnetic moment on the Mn atom. We then define the spin polarization ($\xi$) of the electron conductance as,
\begin{equation}\label{eq1}
\xi=\left|\frac{G^{down}-G^{up}}{G^{up}-G^{down}}\right|
\end{equation}
where G$^{down}$ (G$^{up}$) denotes conductance at Fermi energy for the spin-down (spin-up) channel. The spin polarizaton for S, C, and L doping sites are found to be 92.3\%, 87\%, and 97\%, respectively, suggesting the possibility of using this kind of clusters as spin filter. The high spin polarization originates from the localized magnetic states that are well protected by the ligand shell from the presense of electrodes. As a comparison, we also considered the case of undoped thiolated Au clusters. For undoped clusters, as previously mentioned, the 1 \emph{uB} of magnetic moment is delocalized, spreading over the whole cluster, and when connected to two electrodes, our calculations found that the magnetic moment disappears due to the charge transfer between electrodes and the cluster. As a result, there is no spin polarization in the conductance spectra. 

In conclusion, via first-principles methods, we investigated the geometrical, electronic, magnetic, and spin-dependent transport properties of Mn-doped ligand-protected Au$_{24}$Mn(SH)$_{18}$ clusters. We found that the most favorable doping position for the Mn atom is the surface site rather than the previously suggested center site due to the less structure expansion. Spin-dependent transport calculations showed that these Mn-doped clusters may be used as nearly-perfect spin filters. The spin-filtering originates from the localized magnetic moment of the clusters that is well protected by ligand from the presence of electrodes, and therefore does not require the symmetry breaking of spin channels in electrodes.

This research was supported by NUS Academic Research Fund (C.Z.)
(Grant Nos: R-144-000-237133 and R-144-000-255-112).


\begin{thebibliography}{}

\bibitem{syth}
P. D. Jadzinsky, G. Calero, C. J. Ackerson, D. A. Bushnell, and R. D. Kornberg, Science 318, 430 (2007).
\bibitem{catalysis}
O. Lopez-Acevedo, K. A. Kacprzak, J. Akola, and H. Hakkinen, Nat. Chem. 2, 329 (2010).
\bibitem{sensors}
M. C. Daniel and D. Astruc, Chem. Rev. 104, 293 (2004).
\bibitem{flor}
Z. K. Wu and R. C. Jin, Nano Lett. 10, 2568 (2010).
\bibitem{PNAS}
M. Walter, J. Akola, O. Lopez-Acevedo, P. D. Jadzinsky, G. Calero, C. J. Ackerson, R. L. Whetten, H. Gronbeck, and H. Hakkinen, Proc. Natl. Acad. Sci. U. S. A. 105, 9157 (2008).
\bibitem{Au25_1}
J. Akola, M. Walter, R. L. Whetten, H. Hakkinen, and H. Gronbeck, J. Am. Chem. Soc. 130, 3756 (2008).
\bibitem{Au25_2}
M. W. Heaven, A. Dass, P. S. White, K. M. Holt, and R. W. Murray, J. Am. Chem. Soc. 130, 3754 (2008).
\bibitem{Au24Pd_ex}
C. A. Fields-Zinna, M. C. Crowe, A. Dass, J. E. F. Weaver, and R. W. Murray, Langmuir 25, 7704 (2009).
\bibitem{CuAgCd_dope}
M. Walter and M. Moseler, J. Phys. Chem. C 113, 15834 (2009).
\bibitem{mod3}
D. E. Jiang and R. L. Whetten, Phys. Rev. B 80 115402 (2009).
\bibitem{mod4}
J. U. Reveles, P. A. Clayborne, A. C. Reber, S. N. Khanna, K. Pradhan, P. Sen, and M. R. Pederson, Nat. Chem. 1, 310 (2009).
\bibitem{mole_mag1}
L. Bogani and W. Wernsdorfer, Nat. Mater. 7, 179 (2008).
\bibitem{mole_mag2}
S. A. Wolf, D. D. Awschalom, R. A. Buhrman, J. M. Daughton, S. von Molnar, M. L. Roukes, A. Y. Chtchelkanova, and D. M. Treger, Science 294, 1488 (2001).
\bibitem{Fe_MgO_Fe}
S. Yuasa, T. Nagahama, A. Fukushima, Y. Suzuki, and K. Ando, Nat. Mater. 3, 868 (2004).
\bibitem{VASP}
G. Kresse and J. Hafner, Phys. Rev. B 47, 558 (1993).
\bibitem{PBE}
J. P. Perdew, K. Burke, and M. Ernzerhof, Phys. Rev. Lett. 77, 3865 (1996).
\bibitem{ATK1}
M. Brandbyge, J. L. Mozos, P. Ordejon, J. Taylor, and K. Stokbro, Phys. Rev. B 65 (2002).
\bibitem{ATK1}
M. Brandbyge, J. L. Mozos, P. Ordejon, J. Taylor, and K. Stokbro, Phys. Rev. B 65 (2002).
\bibitem{ATK2}
J. Taylor, H. Guo, and J. Wang, Phys. Rev. B 63 121104(2001).
\bibitem{JPCL}
C. M. Aikens, J. Phys. Chem. Lett. 2, 99 (2011).
\bibitem{neutral}
M. Z. Zhu, C. M. Aikens, M. P. Hendrich, R. Gupta, H. F. Qian, G. C. Schatz, and R. C. Jin, J. Am. Chem. Soc. 131, 2490 (2009).
\bibitem{Au_S1}
G. S. Tulevski, M. B. Myers, M. S. Hybertsen, M. L. Steigerwald, and C. Nuckolls, Science 309, 591 (2005).


\end{thebibliography}
\end{document}